\documentclass[11pt,english]{iopart}
\usepackage{graphicx}
\usepackage{tikz}
\expandafter\let\csname equation*\endcsname\relax 
\expandafter\let\csname endequation*\endcsname\relax
\usepackage[version=3]{mhchem}
\mhchemoptions{arrows=pgf}
\usepackage{babel}
\usepackage{url}
\usepackage[range-phrase = --,range-units = single]{siunitx}
\usepackage{epstopdf}
\begin{document}

\rapid{Exploring the temporally resolved electron density evolution in Extreme Ultra-Violet induced plasmas}

\author{R M van der Horst, J Beckers, S Nijdam and G M W Kroesen}
\ead{r.m.v.d.horst@tue.nl}
\address{Department of Applied Physics, Eindhoven University of Technology, PO Box 513, 5600MB Eindhoven, The Netherlands}

\begin{abstract}
We measured the electron density in an Extreme Ultra-Violet (EUV) induced plasma. This is achieved in a low-pressure argon plasma by using a method called microwave cavity resonance spectroscopy. The measured electron density just after the EUV pulse is \SI{2.6E16}{\per\cubic\meter}. This is in good agreement with a theoretical prediction from photo ionization, which yields a density of \SI{4.5E16}{\per\cubic\meter}. After the EUV pulse the density slightly increase due to electron impact ionization. The plasma (i.e. electron density) decays in tens of microseconds.
\end{abstract}

\submitto{\JPD}

\maketitle

% Introduction
Industries are striving to reduce the size of computer chips to meet the continuous demand of increasing computer speed and memory capacity. This miniaturization can be achieved by reducing the wavelength of the light used in lithography machines. In modern machines, high energetic (\SI{92}{\electronvolt}) photons with wavelengths as short as \SI{13.5}{\nano\meter} are generated by tin-based pulsed plasma sources. These photons are directed through a complex configuration of multilayer mirrors to the desired location in the scanner. Although gas pressures are being kept relatively low (\SIrange{0.1}{30}{\pascal}), these photons ionize the background gas. Consequently, a plasma is induced everywhere the beam of photons travels through the machine. In the multi-billion dollar lithography industry the formation of these plasmas is of significant importance since in the vicinity of surfaces, strong electric fields may be induced by the formation of a plasma sheath. These strong fields accelerate ions up to high velocities towards highly delicate surfaces such as multilayer mirrors. If these ions gain enough energy, they may pose a potential threat to the expensive mirrors. Therefore, characterization of the Extreme Ultra-Violet (EUV) induced  plasma is essential.

Since the plasma is driven by interactions of electrons with background particles, the electron density is one of the most important plasma parameters. In previous studies, simulations have been performed to determine the electron density in these plasmas~\cite{vandervelden2006}. Also, attempts have been made to characterize the plasma with Langmuir probes, however the authors themselves already concluded that these probes are not feasible to determine the electron density and temperature~\cite{vandervelden2008}. We measured the electron density in these EUV induced plasmas.

\begin{figure}
\centering
\includegraphics{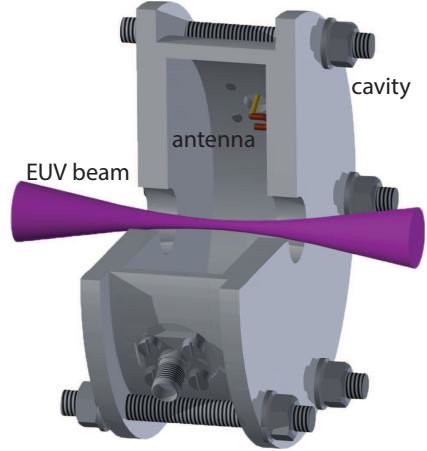}
\caption{Drawing of the microwave cavity with an indication of the EUV beam.}
\label{fig:experimentalSetup}
\end{figure}

% Experimental method
The electron density is determined using a method called microwave cavity resonance spectroscopy (MCRS). This method is previously used in other studies to determine the electron density in radio-frequency (RF) plasmas~\cite{vandewetering2012,beckers2009,stoffels1995} and discharge tubes~\cite{gundermann1999}. Since the MCRS technique is explained in great detail in those publications, we suffice here with a brief description of the key concepts. In MCRS measurements a standing microwave is excited in a cylindrical cavity. The cavity has an entrance and exit hole, so the EUV beam can pass through the cavity (see figure~\ref{fig:experimentalSetup}). However, the size of these holes is kept smaller than the wavelength of the microwave, such that no microwave radiation leaks out of the cavity. The excited standing wave only exists at specific frequencies, the resonant frequencies $\omega_{0}$, which depend, amongst others, on the permittivity of the medium inside the cavity. When free electrons are created inside the cavity by the EUV beam, the permittivity will change and therefore the resonant frequency will shift. The electron density can now be determined by measuring the change in resonant frequency $\Delta\omega$~\cite{vandewetering2012}:
\begin{equation}
\bar{n}_{e}=\frac{2m_{e}\varepsilon_{0}}{e^{2}}\frac{\omega^{2}}{\omega_{0}}\Delta\omega
\end{equation}
with $m_{e}$ the electron mass, $e$ the elementary charge and $\omega_{0}$ and $\omega$ the resonant frequencies without and with plasma, respectively. This equation is only valid if the plasma frequency $\omega_p=\sqrt{n_ee^2/m_e\varepsilon_0}$ is much smaller than $\omega$ and if the collision frequency is much smaller than $\omega$. Note that the number found for $\bar{n}_{e}$ with this method is the volume averaged electron density weighted with the square of the electric field of the standing wave of the mode $\vec{E}$~\cite{vandewetering2012}:
\begin{equation} 
\bar{n}_{e}=\frac{\int_{cavity}n_{e}(\vec{r})E^{2}(\vec{r})d\vec{r}}{\int_{cavity}E^{2}(\vec{r})d\vec{r}}.
\label{eq:avDensity}
\end{equation}
In our experiments we excite the TM$_{010}$ mode, which has the largest electric field near the axis of the cavity. That means that the electron density is mainly sampled at the axis, which is also the position of the plasma. For our measurements we averaged \num{700} EUV pulses. The standard deviation is taken as an indication of the error in the squared-electric-field weighted averaged electron density. This error is less than \SI{30}{\%}.

\begin{figure}
\centering
\includegraphics{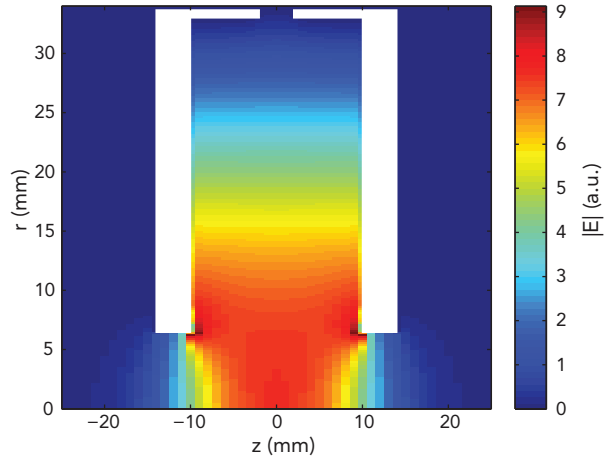}
\caption{Modeled electric field of the TM$_{010}$ mode in our cavity. The white geometry represents the walls of the cavity. The opening at $r=\SI{33}{\milli\meter}$ represents the antenna and the openings at $z=\SI{\pm 10}{\milli\meter}$ represent the entrance and exit holes for the EUV beam. The electric field is calculated with the plasimo platform~\cite{diaz2012,vandijk2009}.}
\label{fig:eField}
\end{figure}

% Experimental setup
The cavity that is used has a radius of \SI{33}{\milli\meter} and a high of \SI{20}{\milli\meter}. The entrance and exit holes of the cavity both have a radius of \SI{6.5}{\milli\meter}. It is made of aluminium. Inside the cavity two copper antennas are mounted at opposite sides, one transmitter and one receiver. The position and shape of the antennas is chosen such that the TM$_{010}$ mode can be excited. We excite this mode with a low-power (\SI{10}{\milli\watt}) microwave generator. The measured frequency of this mode was \SI{3.49674+-0.00004}{\giga\hertz}. Therefore we have a lower detection limit of \SI{1E12}{\per\cubic\meter} and an upper detection limit of \SI{1E17}{\per\cubic\meter}. Figure~\ref{fig:eField} shows the electric field calculated with the plasimo platform~\cite{diaz2012,vandijk2009}. Since the electric field is largest at the axis, the electron density in mainly sampled there (see equation~\ref{eq:avDensity}). The response of the cavity is measured with a microwave detector with a time response of \SI{10}{\nano\second}. The resonant frequency is determined by sweeping the frequency of the microwave generator and measuring the cavity response at every frequency. The frequency at which the cavity response is maximum, is the resonant frequency. The full width at half maximum (FWHM) $\Gamma$ of the resonant peak is used to determine the quality factor of the cavity: $Q=\omega_0/\Gamma=\num{150}$. This gives a response time of the cavity ($\tau=2Q/\omega_0$) of  \SI{14}{\nano\second}. The EUV radiation is produced with a xenon pinch plasma delivering an energy density of \SI{4.8}{\joule\per\square\meter} per pulse with a duration of about \SI{150}{\nano\second}. The radiation is focused in the centre of the cavity. The FWHM of the beam is there about \SI{4}{\milli\meter} and the divergence is about \SI{10}{\degree}. A spectral purity filter (SPF) is placed between the source and the cavity. This SPF is made of a \SI{200}{\nano\meter} foil that transmits between approximately \SI{10}{\nano\meter} and \SI{20}{\nano\meter}. The pressure is measured with a \SI{0.1}{\milli\bar} Baratron capacitive gauge.

\begin{figure}
\centering
\includegraphics{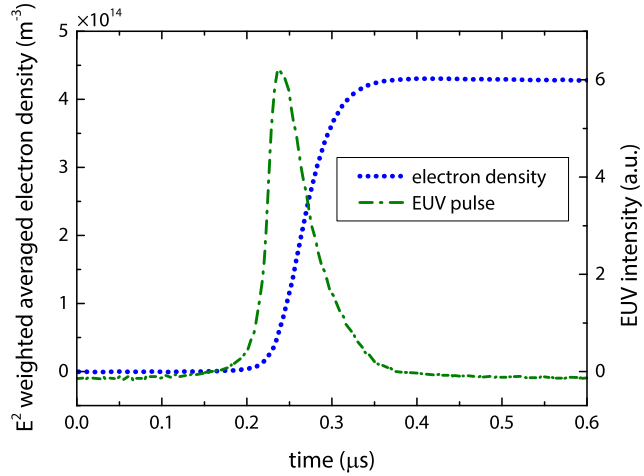}
\caption{Electron density and relative EUV intensity as function of time during the EUV pulse in an EUV induced plasma in \SI{2.5}{\pascal} argon. The signals are corrected for the delay in the cables, the uncertainty is about \SI{20}{\nano\second}. Figure~\ref{fig:electronDensity_1} shows the same measurement on a larger timescale.}
\label{fig:electronDensity_2}
\end{figure}

\begin{figure}
\centering
\includegraphics{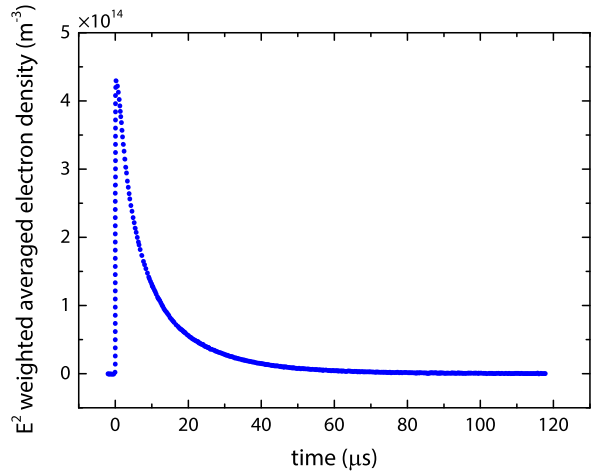}
\caption{Electron density as function of time in an EUV induced plasma in \SI{2.5}{\pascal} argon on a larger timescale.}
\label{fig:electronDensity_1}
\end{figure}

% Results
The measured electron density in an EUV induced plasma in argon at \SI{2.5}{\pascal} is shown in figures~\ref{fig:electronDensity_2}~and~\ref{fig:electronDensity_1}. It should be noted that the density in these graphs is the squared-electric-field weighted averaged electron density (see equation~\ref{eq:avDensity}). At the start of the EUV pulse, the electron density starts to increase. At the end of the EUV pulse, the density still increases slightly to a maximum value of \SI{4.4e14}{\per\cubic\meter}. After the maximum density is reached, the plasma decays in tens of microseconds. We will discuss the temporal behaviour in the next paragraphs.

The initial increase in electron density at the start of the EUV pulse is caused by absorption of EUV light and subsequent photo-ionization:
\begin{equation}
\ce{\gamma\, + Ar -> e- + Ar+}.
\end{equation}
The ionization energy for argon is \SI{15.8}{\electronvolt}~\cite{kramida2013}. Almost all excess energy (\SI{76}{\electronvolt}) is transferred to the ejected electron due to momentum conversation. The ions will remain approximately at room temperature. The maximum electron density due to photo-ionization can be estimated from the absorbed EUV energy $I_{abs}(\lambda)$:
\begin{equation}
n_{e}  =  \int_{0}^{\infty}\frac{I_{abs}(\lambda)}{L_{c}hc/\lambda}d\lambda,
\end{equation}
with $I_{abs}(\lambda)$ the absorbed EUV irradiation as function of wavelength $\lambda$, $L_{c}$ the length of the cavity, $h$ the Planck constant and $c$ the speed of light. Using Beer-Lambert law, this can be rewritten:
\begin{equation}
n_{e}=\frac{1}{L_{c}hc}\int_{0}^{\infty}I_{0}(\lambda)
\left[1-\exp\left\{ -n_{Ar}\sigma(\lambda)L_{c}\right\} \right]
\lambda d\lambda,
\end{equation}
with $I_{0}(\lambda)$ the EUV irradiation as function of wavelength $\lambda$ before the cavity, $n_{Ar}$ the background argon density and $\sigma(\lambda)$ the photo-ionization cross section. The photo-ionization cross section is taken from Marr and West~\cite{marr1976}. The EUV irradiation $I_{0}(\lambda)$ is calculated from the relative spectrum recorded by Kieft~\cite{kieft2008} and a wavelength integrated power measurement of the EUV source (\SI{4.8}{\joule\per\square\meter} per pulse). This results in an electron density of \SI{4.5e16}{\per\cubic\meter}. To compare this to the experimental results, we need to assume a spatial profile for the electron density. Assume that the spatial profile of the electron density during the EUV pulse is Gaussian with a FWHM of \SI{4}{\milli\meter} equal to the FWHM of the EUV beam. Since we also know the mode of the cavity (TM$_{010}$), we can use equation~\ref{eq:avDensity} to convert the averaged electron density ($\bar{n}_{e}=\SI{4E14}{\per\cubic\meter}$) to a density on the axis of the cavity. This results in an electron density on the axis at the end of the EUV pulse of about \SI{2.6E16}{\per\cubic\meter}. This is in good correspondence with the estimation from the power absorption (\SI{4.5e16}{\per\cubic\meter}).

The electron induced ionization cross section for the high energy (92\,eV) electrons in argon is $\sigma_{ea}=\SI{2.8E-20}{\square\meter}$~\cite{phelps1999}. For the typical pressures used here (\SI{2.5}{\pascal}), the mean free path of these high energy electrons is then about \SI{6}{\centi\meter}, so the first of these electrons will reach the wall of the cavity in a few nanoseconds before interacting with the background gas. The slower argon ions remain on the axis at the position they have been created. Consequently, a potential difference is generated between the plasma and the wall. This potential traps the remaining electrons in the cavity. When these trapped electrons collide with an argon atom, the atom will be ionized, thereby creating a secondary electron:
\begin{equation}
\ce{e- + Ar -> e- + e- + Ar+}.
\end{equation}
Due to this process the electron density still increases after the EUV pulse.

The potential decelerates the electrons and accelerates the ions to the wall. Therefore the electrons and ions will diffuse at the same speed. When they reach the wall of the cavity, they will recombine.

% Conclusion
In conclusion, we measured the electron density in an EUV induced plasma with microwave cavity resonance spectroscopy. Electrons are created via two processes. First electrons are generated by photo-ionization. These high energy electrons create secondary electrons via collisions with the background gas. The electron density at the end of the EUV pulse is in good agreement with a theoretical prediction.

\ack
The authors would like to acknowledge ASML for their financial support and the opportunity to use its EUV sources. We would like to thank V. Banine, E. Osorio and M. van Kampen for the fruitful discussions.

\section*{References}

\bibliographystyle{iopart-num}
\bibliography{references}

\end{document}